\documentclass[onecollarge]{svjour2}

\smartqed

\usepackage{graphicx}

\journalname{Acta Mechanica}

\begin{document}

\title{Phase Slip Solutions in Magnetically Modulated Taylor-Couette Flow}
\titlerunning{Magnetically Modulated Taylor-Couette Flow}

\author{Rainer Hollerbach \and Farzana Khan}

\institute{R. Hollerbach \at
Department of Applied Mathematics, University of Leeds, Leeds, LS2 9JT, UK\\
              \email{rh@maths.leeds.ac.uk}
\and       F. Khan \at
Department of Applied Mathematics, University of Leeds, Leeds, LS2 9JT, UK\\
\emph{Permanent address:} Department of Mathematics, Quaid-i-Azam University,
Islamabad, Pakistan}

\date{Received: date / Accepted: date}

\maketitle

\begin{abstract}
We numerically investigate Taylor-Couette flow in a wide-gap configuration,
with $r_i/r_o=1/2$, the inner cylinder rotating, and the outer cylinder
stationary.  The fluid is taken to be electrically conducting, and a magnetic
field of the form $B_z\approx(1 + \cos(2\pi z/z_0))/2$ is externally imposed,
where the wavelength $z_0=50(r_o-r_i)$.  Taylor vortices form where the field
is weak, but not where it is strong.  As the Reynolds number measuring the
rotation rate is increased, the initial onset of vortices involves phase slip
events, whereby pairs of Taylor vortices are periodically formed and then
drift outward, away from the midplane where $B_z=0$.  Subsequent bifurcations
lead to a variety of other solutions, including ones both symmetric and
asymmetric about the midplane.  For even larger Reynolds numbers a different
type of phase slip arises, in which vortices form at the outer edges of the
pattern and drift inward, disappearing abruptly at a certain point.  These
solutions can also be symmetric or asymmetric about the midplane, and
co-exist at the same Reynolds number.  Many of the dynamics of these phase
slip solutions are qualitatively similar to previous results in geometrically
ramped Taylor-Couette flows.
\end{abstract}

\section{Introduction}

The flow between differentially rotating cylinders, known as Taylor-Couette
flow (TC flow), is one of the oldest problems in fluid dynamics, but
continues to attract considerable experimental \cite{vG11,AH13,CC14} as well
as numerical \cite{YWB05,WT12,S15} interest.  Another indication of its
continuing relevance is the number of distinct branches it has spawned.  Two
of these are: (a) ramped TC flow, where the inner and/or outer radii of the
cylinders are not uniform, but vary along the axial direction, and (b) magnetic
TC flow, where the fluid is taken to be electrically conducting, and magnetic
fields are externally applied.  In this work we combine these two branches, by
using magnetic fields to impose the axial modulation that is otherwise created
by the geometrical ramping.  We demonstrate that the phenomenon of phase slip,
whereby pairs of Taylor vortices are either created or destroyed, carries over
from geometrically to magnetically ramped TC flows.  We suggest that
magnetically modulated TC flow may be a more convenient system to numerically
study some of the resulting pattern formation effects.

The configuration of greatest interest in ramped TC flow is when the radii of
the cylinders gradually vary in such a way that parts of the domain can be
above the critical Reynolds number for the onset of Taylor vortices while
other parts are still below.  Refs.~\cite{PM79,K82} considered such
subcritical ramps in a general pattern formation context, and showed that the
usual Eckhaus-stable wave-number band instead collapses to a single
wave-number.  Riecke \& Paap \cite{RP87} applied this so-called
phase-diffusion equation approach specifically to TC flows, and demonstrated
that ``there exist ramps which do not permit any static patterns but force them
to drift.''  That is, once the wave-number is fixed, at the transition point(s)
from locally subcritical to supercritical, that can force the wave-number in
other parts of the supercritical region to be Eckhaus-unstable.  The pattern
responds with a phase slip event, essentially an attempt to move the
wave-number toward Eckhaus stability.  The resulting Taylor vortices drift
away though, forcing a new phase slip, and so on indefinitely.  (Note however
that not all subcritically ramped systems necessarily lead to phase slip
behaviour; for reviews of various pattern-forming systems without phase slip
see \cite{C05}.)

Occurring in parallel with the theory of Riecke \& Paap were a series of
experiments \cite{C83,D86,N90} which confirmed many of these ideas, including
the uniqueness of the selected wave-number and the possibility of drifting
patterns.  Paap \& Riecke \cite{PR91} in turn followed up with further
work demonstrating that the phase equation approach yields quantitatively
accurate agreement with many of these experimental results.  They further
suggested that it should be possible to construct ramps where the phase slip
events occur irregularly, resulting in spatiotemporal chaos in the pattern.
Ref.~\cite{W97} took up this challenge experimentally, and succeeded in
obtaining a period-doubling cascade to chaotic phase dynamics.  We therefore
conclude our overview of ramped TC flow by noting that there is a broad
variety of possible patterns, and a good theoretical framework for
understanding many of these results.

Turning next to magnetic TC flow, the current focus of attention is primarily
on the magnetorotational instability (MRI), whereby a magnetic field can
destabilize a rotation profile that would otherwise be stable according to
the Rayleigh criterion.  First discovered in 1959 in TC flows \cite{V59},
the MRI lay largely dormant until it was suggested that it might play a
crucial role in astrophysical accretion disks \cite{BH91}.  This discovery
reignited interest in the MRI in TC flows, and specifically the possibility
of obtaining it \cite{J01,RZ01} or variants of it \cite{HR05,HTR10}
experimentally.  The standard MRI has not yet been obtained \cite{N10,R12}, but
the helical \cite{S06} and azimuthal \cite{S14} variants have.  A further recent
magnetic TC experiment \cite{C11} has measured the so-called $\omega$-effect.

However, here we wish to return to some of the earliest
work \cite{C53,DO60,DO62} on magnetic TC flows, dating back over 50 years.
In the standard configuration where the outer cylinder is stationary (and the
Rayleigh criterion therefore does not enforce hydrodynamic stability), both
theory \cite{C53} and experiment \cite{DO60,DO62} agree that imposing a uniform
axial magnetic field has a stabilizing influence, that is, delays the onset of
Taylor vortices to greater Reynolds numbers.  It is this stabilizing feature
that forms the basis of our work here.  In particular, suppose one were to
impose not a uniform axial field, but instead $B_z\approx
(1 + \cos(2\pi z/z_0))/2$ [the precise form is given below in Eq.\ (3)].
That is, the field reaches a maximum at integer multiples of the basic
periodicity $z_0$, and drops to zero at half-integer multiples of $z_0$.
We would then expect to obtain Taylor vortices where the field is weak, and
no Taylor vortices where it is strong -- the same subcritical ramping effect
as before in the geometrically ramped problem.

\section{Equations}

We consider the standard wide-gap Taylor-Couette configuration with radii
$r_i$ and $r_o$ satisfying $r_i/r_o=1/2$.  The inner cylinder rotates at a
rate $\Omega$, and the outer cylinder is stationary.  In the so-called
inductionless limit, the suitably scaled Navier-Stokes and magnetic
induction equations become
$$Re\frac{\partial\bf U}{\partial t}=-\nabla p + \nabla^2{\bf U}
  - Re{\bf U\cdot\nabla U} + Ha^2(\nabla\times{\bf b})\times{\bf B}_0,
\eqno(1)$$
$$\nabla^2{\bf b}=-\nabla\times({\bf U\times B}_0).\eqno(2)$$
Length has been scaled by $r_i$, time by $\Omega^{-1}$, and $\bf U$ by
$\Omega r_i$.  ${\bf B}_0$ is the externally imposed magnetic field, and
$\bf b$ the induced field.  The two nondimensional parameters in these
equations are the usual Reynolds number
$$Re=\frac{\Omega r_i^2}{\nu}$$
measuring the inner cylinder's rotation rate, and the Hartmann number
$$Ha=\frac{B_0 r_i}{\sqrt{\mu\rho\nu\eta}}$$
measuring the strength of the imposed field.  The quantities $\mu$, $\rho$,
$\nu$, and $\eta$ are the fluid's permeability, density, viscosity, and
magnetic diffusivity, respectively.  See also \cite{H09,T11} for a more
detailed derivation of these equations in the inductionless limit, in the
context of magnetic spherical Couette flow.  For experiments in magnetic
spherical Couette flow see \cite{C14,Z14} and further references therein.

The spatial structure of the imposed field ${\bf B}_0$ is given by
$${\bf B}_0 =\bigl[(1 + \cos(\kappa z)I_0(\kappa r))/2\bigr]\,{\bf\hat e}_z
  + \bigl[\sin(\kappa z)I_1(\kappa r)/2\bigr]\,{\bf\hat e}_r,\eqno(3)$$
where $\kappa=2\pi/z_0$, and $I_0$ and $I_1$ are the modified Bessel
functions \cite{AS68}.  The wavelength $z_0$ is an adjustable parameter, but
once fixed, the rest of the structure is completely determined by the
requirements that ${\bf B}_0$ be a potential field, and imposed from the
region $r>r_o$ rather than $r<r_i$.  That is, with a suitable array of
external Helmholtz coils one could actually impose such a field, a point we
will return to in the conclusion.  Note finally that the $r$-dependent parts
of ${\bf B}_0$ are necessary to satisfy $\nabla\cdot{\bf B}_0=0$ and
$\nabla\times{\bf B}_0={\bf0}$, but are in fact quite small.  Using the
asymptotic properties $I_0(\kappa r)\approx1$ and $I_1(\kappa r)\ll1$ for
$\kappa r\ll1$, one obtains ${\bf B}_0\approx[(1+\cos(\kappa z))/2]\,
{\bf\hat e}_z$, as desired to create the magnetic ramping effect.

These equations (1-3), together with boundary conditions no-slip for $\bf U$
and perfectly conducting for $\bf b$, were numerically solved using an
axisymmetric, pseudo-spectral code \cite{H08}.  Very briefly, $\bf U$ and
$\bf b$ are expanded as
$${\bf U}=\nabla\times(\psi\,{\bf\hat e}_\phi) + v\,{\bf\hat e}_\phi,\qquad
  {\bf b}=\nabla\times(a\,{\bf\hat e}_\phi) + b\,{\bf\hat e}_\phi,$$
then $\psi$, $v$, $a$ and $b$ are further expanded in terms of Chebyshev
polynomials in $r$ and Fourier series in $z$.  Typical resolutions used were
$20-30$ Chebyshev polynomials and $200-300$ Fourier modes.  The time-stepping
of Eq.\ (1) is second-order Runge-Kutta, modified to treat the diffusive
terms implicitly.  Eq.\ (2) is directly inverted for $\bf b$ at each
time-step of Eq.\ (1).  Typical time-steps used were $0.02-0.05$.

After preliminary scans in the range $z_0=20-80$ yielded qualitatively
similar phase slip solutions, the axial length was fixed at $z_0=50$.
For comparison, the spatially ramped experiment \cite{W97} most
closely related to our magnetic case had a nondimensional length of 29.
Again after some preliminary scans, it was found that Hartmann numbers in
the range $2-10$ also yielded similar solutions, so only $Ha=5$ was
investigated in further detail.  The single remaining parameter, the Reynolds
number, was then varied throughout the interval $Re=67-85$.  The reason for
the lower limit is simple: in the non-magnetic problem the critical Reynolds
number for the onset of Taylor vortices is known \cite{D84} to be 68.2, so one
would hardly expect anything interesting to happen before then in this problem.
The upper limit was chosen partly because enough interesting things had already
happened by then, and partly because the calculations become more
time-consuming beyond that point.  Eventually of course one would also expect
the solutions to become three-dimensional, although in the non-magnetic problem
at least non-axisymmetric instabilities do not arise until much larger Reynolds
numbers, around three times supercritical \cite{J85} versus less than 30\%
supercriticality considered here.

Finally, it is worth noting that at least some of the cases required extremely
long integration times before periodic solutions emerged.  This is a natural
consequence of having $z_0\gg1$: the diffusive time-scale between two points
separated by the maximum possible distance $z_0/2$ is $Re(z_0/2)^2=O(10^5)$,
so very long time-scales are almost inevitable.

\section{Results}

Figure 1 shows the flow at $Re=67$, still below the onset of Taylor vortices.
There is in fact already a deviation from the ideal Couette profile, most
noticeably in the meridional circulation $\psi$, which is identically zero in
the ideal basic state, but now consists of four large circulation cells
centered on the midplane $z=25$.  These cells already appear for all $Re>0$,
and are analogous to the Ekman cells obtained in cylinders with top and
bottom endplates \cite{D86,G82}.  In this case they are
caused by the $r$-dependent parts of ${\bf B}_0$; if the imposed field were
exactly $B_z=(1+\cos(\kappa z))/2$, then the ideal Couette profile
$v=C_1r+C_2/r$ would also be an exact solution of the Navier-Stokes equation,
along with $\psi=0$.  By taking $z_0$ to be sufficiently large, these
deviations from the ideal profile can thus be made as small as desired.
Geometrically ramped TC flows also have deviations from the ideal Couette
profile, caused by similar dynamics as the endplate-induced Ekman cells.
These deviations from the ideal profile are not crucial though to any of the
phase equation results, including the existence of drifting
patterns \cite{RP87}.

\begin{figure}
  \centerline{\includegraphics{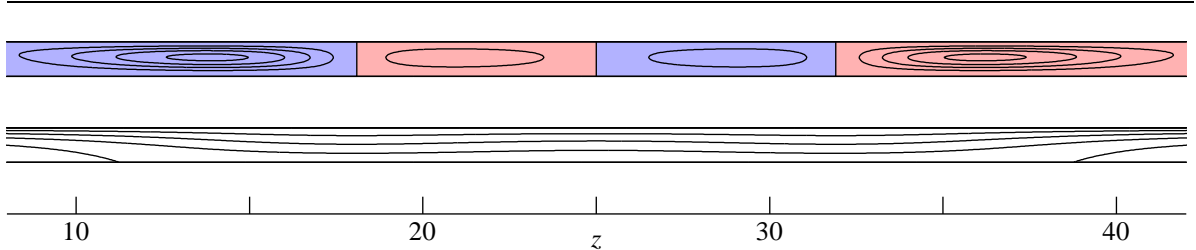}}
  \caption{The solution at $Re=67$.  The top row shows the meridional
circulation $\psi$, with a contour interval of $10^{-3}$, and blue/red
indicating clockwise/counter-clockwise circulation.  The second row shows
$v$, with a contour interval of 0.25.  The horizontal axis corresponds to
length $z$ along the cylinders; the regions $z<8$ and $z>42$ are not shown
as all the relevant dynamics occur in the middle region.  On the vertical
axis $r_i$ is on the top and $r_o$ is on the bottom.}
\label{fig1}
\end{figure}

\begin{figure}
  \centerline{\includegraphics{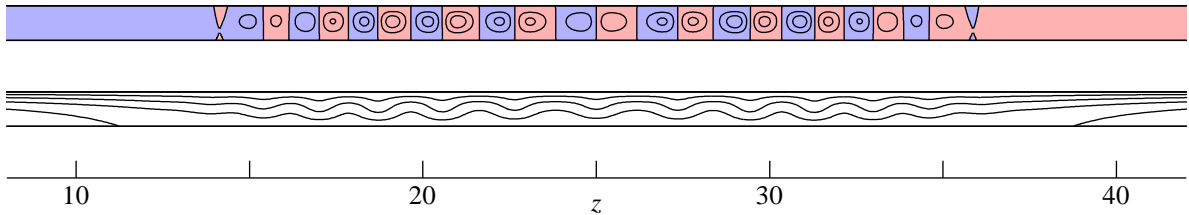}}
  \caption{The solution at $Re=77$.  The contour interval for $\psi$ is now
0.01; everything else is as in figure 1.  Note the Taylor vortices between
$z\approx14$ and 36, in the region where the field is weak.  Based on the
local value of the field, the flow should be linearly unstable to Taylor
vortices in the region $13<z<37$, in excellent agreement.}
\label{fig2}
\end{figure}

Figure 2 shows the flow at $Re=77$.  As expected, we find Taylor vortices in
the middle, where the imposed field is weak, but not near the ends, where it
is strong.  This solution is in fact steady, without either drift or phase
slip events.  However, it is also not the initial onset of Taylor vortices;
instead, these solutions exist only in the range $76.5< Re\le81.3$.  Note how
much stronger the meridional circulation in the Taylor vortices is, in
comparison with the background cells from figure 1.  This background
circulation is still present even here, but is overwhelmed by the Taylor
vortices, with the result that clockwise and counter-clockwise vortices are
almost the same strength.

If figure 2 does not represent the initial onset, then at what Reynolds number
do the Taylor vortices first arise, and how?  The bifurcation point occurs at
$Re=67.4$, via a supercritical Hopf bifurcation, that is, a drifting pattern.
Two further comments are also in order regarding this initial bifurcation.
First, the slight reduction from 68.2 in the non-magnetic case \cite{D84} to
67.4 here is caused by the presence of the background cells, and does not
indicate a subcritical bifurcation.  Second, as a transition from a steady
to a periodic solution, it is a true bifurcation.  This is different from TC
flows with endplates, where the background Ekman cells cause the bifurcation
to be imperfect \cite{RC04}, from one steady pattern to another, but without
any true bifurcation.

To best illustrate the full time-dependence of the drifting pattern, including
the phase slip events that form a crucial part of it, we begin by noting that
the radial structure of the vortices is relatively straightforward: $\psi=0$
at either boundary, and in between is positive/negative for
clockwise/counter-clockwise vortices.  Simply focusing on the middle $r=1.5$
will therefore capture all the essential features.  Contour plots of
$\psi(t,z,1.5)$ will then reveal the structure in both time and length along
the cylinder.

As illustrated in figure 3, the solution consists of a series of phase slips
at the midplane $z=25$.  The newly created Taylor vortices drift outward,
eventually fading away at $|z-25|\approx10$, where the magnetic field becomes
too strong for them to persist.  Note also the slight asymmetry between the
two phase slip events that constitute one period of the pattern.  This is
again caused by the weak background circulation; newly created vortices
with circulation in the opposite sense as the background (figure 1) persist
slightly longer than vortices with circulation in the same sense.  The period
of these solutions gradually increases from $T=1243$ at $Re=67.4$ to $T=4393$
at $Re=74.5$, the value shown in figure 3.

\begin{figure}
  \centerline{\includegraphics{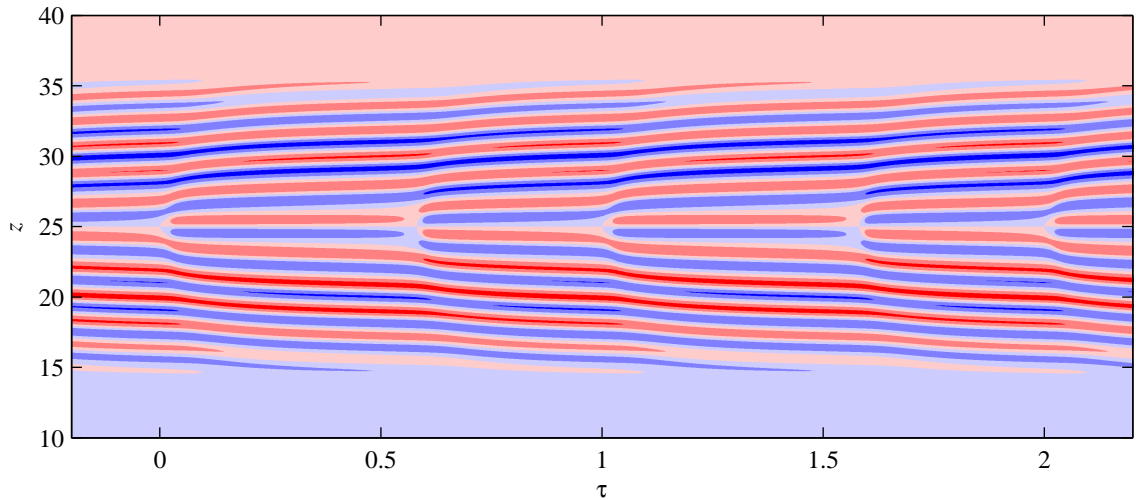}}
  \caption{Contour plots of $\psi(t,z,1.5)$ at $Re=74.5$, with a contour
interval of 0.01.  Blue/red represents clockwise/counter-clockwise Taylor
vortices.  On the horizontal axis $\tau=t/T$, where the period $T=4393$ at
this Reynolds number.  Note the midplane symmetry $\psi(t,z)=-\psi(t,50-z)$,
and the slightly uneven timing between the two phase slips per period.}
\label{fig3}
\end{figure}

The next two bifurcations occur at $Re=74.8$ and $Re=75.8$.  The first one
breaks the midplane symmetry $\psi(t,z)=-\psi(t,50-z)$ seen in figure 3, but
still preserves the shift-and-reflect symmetry $\psi(t,z)=-\psi(t+T/2,50-z)$.
Correspondingly, the average value of the asymmetric component over a period
is still zero.  The second one breaks the shift-and-reflect symmetry as well.
There are thus two solutions, with average asymmetric components of either
sign (just as in a pitchfork bifurcation).  Figures 4 and 5 show examples of
these solutions, at $Re=75.5$ and $Re=76$.  The first bifurcation already
causes the period to double, since it now takes two of the original cycles for
the asymmetry to occur first in one half and then in the other.  Beyond that,
the periods still continue increasing, from $T=13334$ at $Re=75.5$ to
$T=22616$ at $Re=76$.  Both of these symmetry-breaking bifurcations are also
supercritical, with no hysteresis if $Re$ is reduced again.

\begin{figure}
  \centerline{\includegraphics{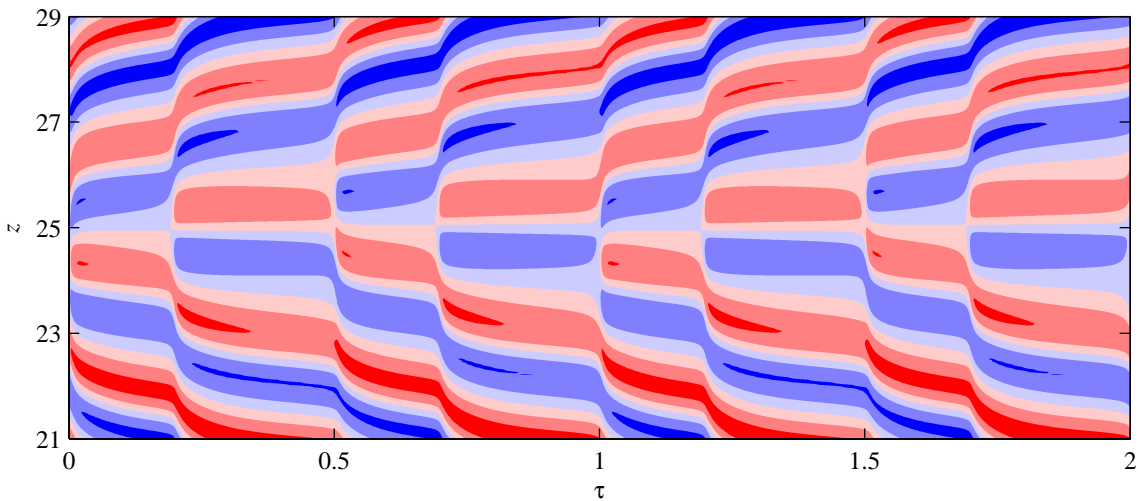}}
  \caption{As in figure 3, but at $Re=75.5$, where the period $T=13334$.
Note the shift-and-reflect symmetry $\psi(t,z)=-\psi(t+T/2,50-z)$, and the
four phase slips per period.}
\label{fig4}
\end{figure}

\begin{figure}
  \centerline{\includegraphics{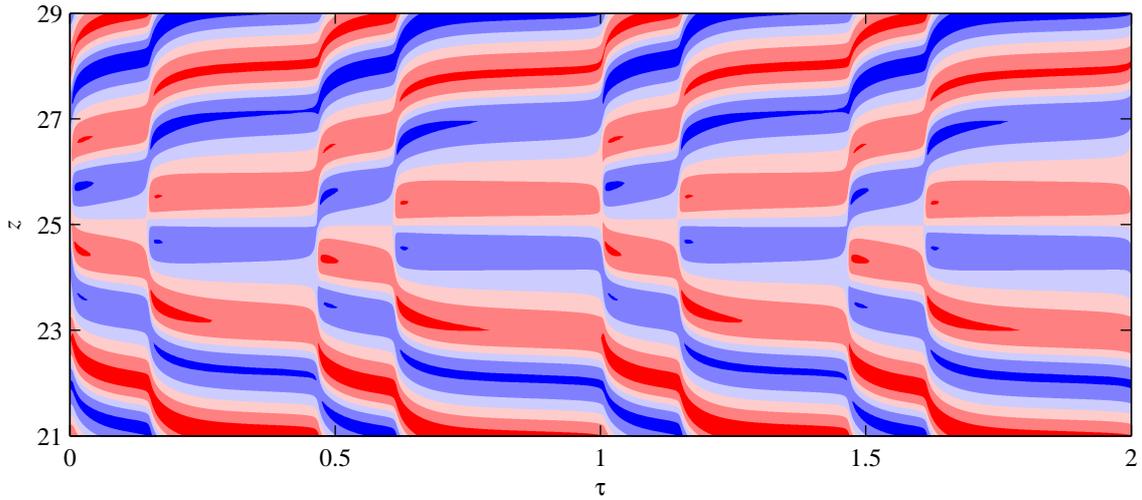}}
  \caption{As in figure 3, but at $Re=76$, where the period $T=22616$.  Note
the four phase slips per period, and the uneven timing between successive
pairs; this is the most obvious manifestation of the loss of the previous
shift-and-reflect symmetry in figure 4.}
\label{fig5}
\end{figure}

Figure 6 shows the pattern at $Re=76.5$.  We notice two differences in
comparison with figure 5.  First, the asymmetry is considerably greater,
with the pattern shifted so much in $z$ that there is now a Taylor vortex
sitting right on the midpoint $z=25$.  Figure 7 compares the entire range
$67\le Re\le 77$, and quantifies items such as how the strength of the Taylor
vortices gradually increases, how the degree of asymmetry increases, and how
the period varies.

\begin{figure}
  \centerline{\includegraphics{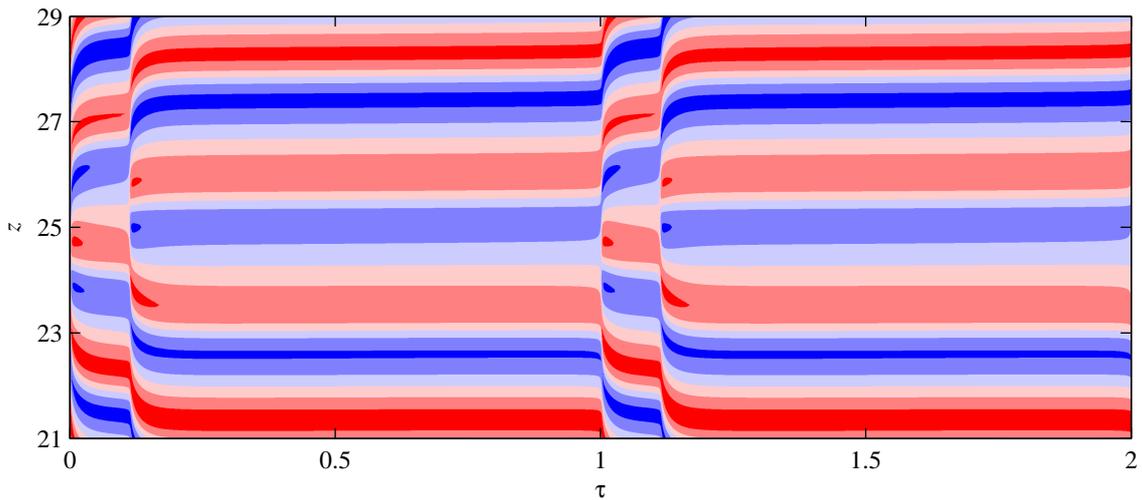}}
  \caption{As in figure 3, but at $Re=76.5$, where the period $T=54388$.
Note the two phase slips per period, and the extreme disparity in timing
between the two, with the pattern almost stationary for most of the cycle.}
\label{fig6}
\end{figure}

\begin{figure}
  \centerline{\includegraphics{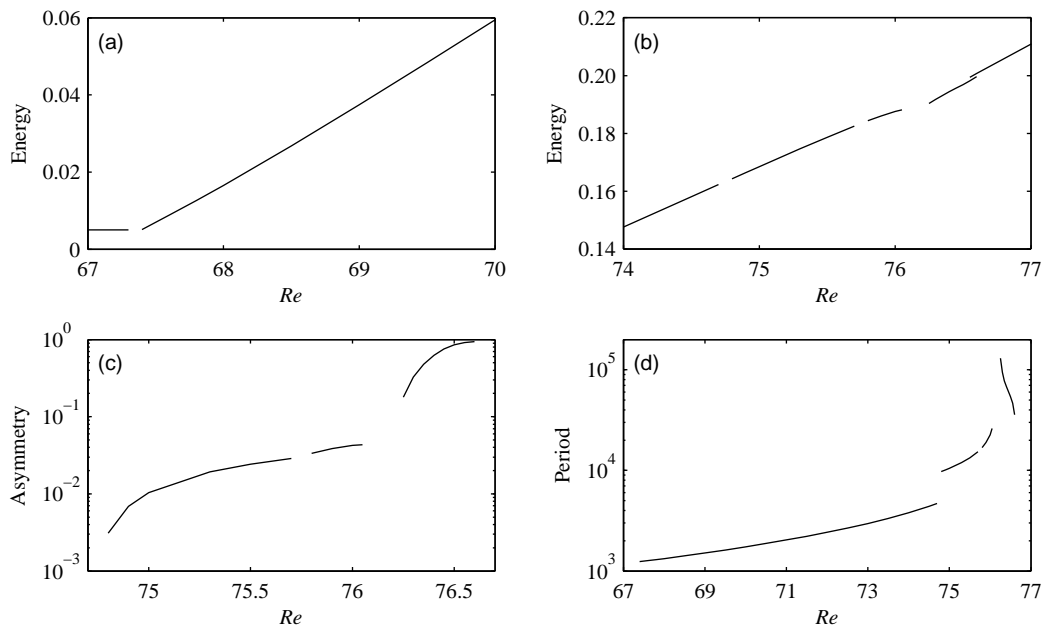}}
  \caption{Panels (a) and (b) both show the kinetic energy in the meridional
circulation (period-averaged for the time-dependent solutions).  Note how the
energy is essentially constant before the onset of Taylor vortices, and
increases linearly thereafter (including in the gap $70<Re<74$ between the
two panels).  Panel (c) shows the fraction of the energy that is contained in
the asymmetric component.  Note how it increases from 0 for $Re<74.8$ toward
almost 1 at the turning point $Re=76.6$.  Panel (d) shows the period; note the
extreme variation over more than two orders of magnitude.  Different line
segments correspond to the different types of solutions discussed in the text.}
\label{fig7}
\end{figure}

The second difference is that the pattern in figure 6 has only two phase slip
events per period, whereas in figure 5 there are four per period.  The most
natural way to connect the two therefore would be to have a period-doubling
bifurcation as $Re$ is reduced from 76.5 back toward 76.  Pinning down
exactly where this bifurcation occurs was unfortunately not possible.  For
$Re\le76.05$ the solutions have four phase slips per period, and for
$Re\ge76.25$ they have two per period.  However, between 76.05 and 76.25 they
not only have very long cycle times (see figure 7), but even after integrating
to $t=10^6$ the solutions still had not settled in to a precise periodicity.
It is possible of course that the solutions in this narrow gap really are
chaotic, and bifurcate to a four-cycle at one end and a two-cycle at the other.

Increasing $Re$ even further, the pattern in figure 6 persists up to
$Re=76.63$, where a turning point is reached, and the solution collapses back
to the steady, symmetric state as in figure 2.  Reducing $Re$ again, the
figure 2 solutions become unstable to a subcritical pitchfork bifurcation at
$Re=76.53$.  There is thus a small but measurable degree of hysteresis in this
transition.  Somewhere on the unstable branch between 76.53 and 76.63 the
solution presumably also undergoes a Hopf bifurcation, thereby acquiring the
time-dependence that it has in the figure 6 solutions.

Finally, on the steady, symmetric figure 2 branch, what happens if $Re$ is
increased rather than decreased?  At $Re=81.4$ a Hopf bifurcation occurs, or
rather two essentially simultaneously, corresponding to symmetric and
asymmetric perturbations.  As a result, attempting to unravel the full details
of the bifurcation diagram proved fruitless.  The underlying dynamics are
quite straightforward though.  Figures 8 and 9 show two solutions that both
exist at $Re=85$.  We note first that they are very different from the
previous phase slip solutions.  Rather than having pairs of vortices created
in the middle and then drifting outward, we now have vortices moving in from
the edges, and then being abruptly destroyed when $|z-25|$ is around 8 or 9.
Based on the local value of the field, the flow should be linearly unstable
to Taylor vortices in the region $11<z<39$.  All of these dynamics are
thus happening within the expected region, but it is not clear what singles
out the particular location where the phase slips occur.

\begin{figure}
  \centerline{\includegraphics{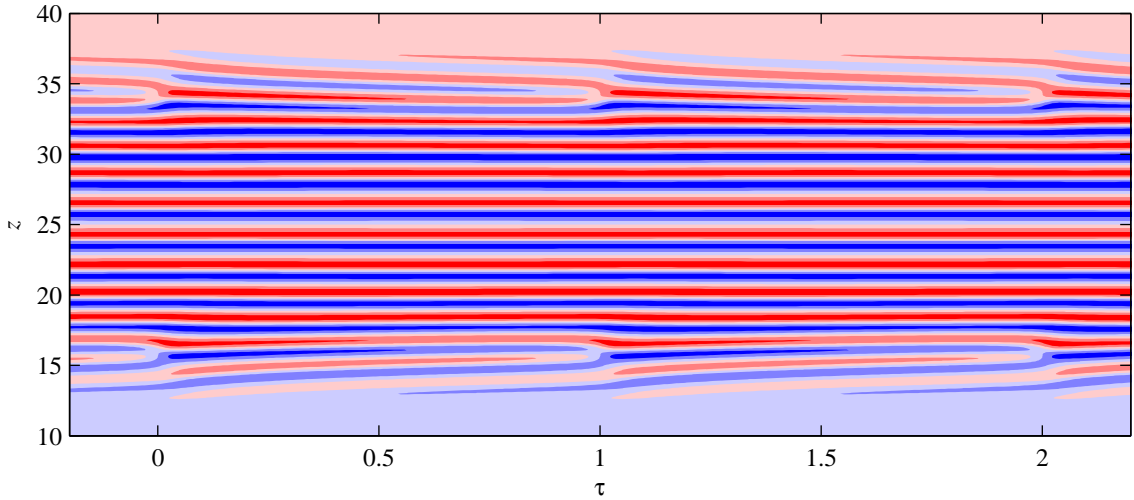}}
  \caption{One possible solution at $Re=85$, with $T=638$.  Note the midplane
symmetry, and the 18 essentially steady Taylor vortices separating the two
time-dependent regions.  The contour interval is 0.012.}
\label{fig8}
\end{figure}

\begin{figure}
  \centerline{\includegraphics{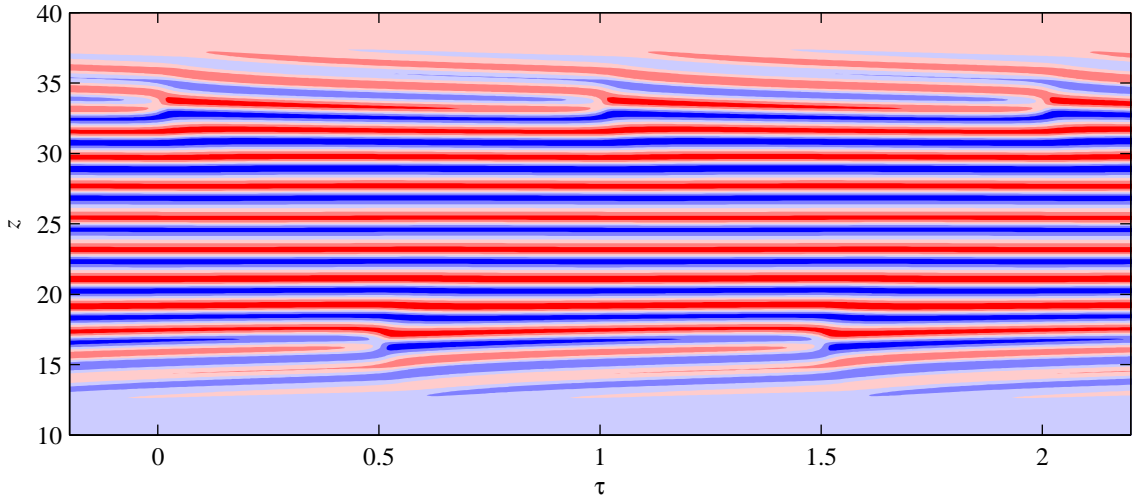}}
  \caption{Another solution at $Re=85$, with $T=755$.  Note the
shift-and-reflect symmetry, and the 16 essentially steady Taylor vortices
separating the two time-dependent regions.  The contour interval is 0.012.}
\label{fig9}
\end{figure}

Comparing figures 8 and 9, the most obvious difference between them is that
figure 8 is symmetric, whereas figure 9 is asymmetric.  That is, in figure 8
the phase slips in the top and bottom halves are in phase in time, whereas in
figure 9 they are exactly half a period out of phase.  This pattern presumably
also explains why the two perturbation types both arose at virtually the same
critical Reynolds number: If the two time-dependent regions are separated in
space by such a large region that is essentially stationary, then the coupling
between them is extremely weak, so there is almost nothing to fix their
relative phases in time.  This very weak coupling between the two regions also
means that very long runs were required in some cases before a definite phase
relationship emerged.  Indeed, there may also be some Reynolds numbers where
no definite relationship ever arises, resulting in some type of quasi-periodic
solution.  It is difficulties like these that lead us to concentrate on the
single value $Re=85$, rather than attempt to map out a full bifurcation
diagram over the range $Re\ge81.4$.

One further, not quite so obvious difference between figures 8 and 9 is
revealed by counting the number of vortices in the stationary region, between
the phase slip events.  In figure 8 there are 18, in figure 9 only 16.  The
easiest way to spot that the number is certainly different is simply to note
that the coloring just above/below $z=25$ is blue/red in figure 8, but
red/blue in figure 9.  These particular solutions were obtained by starting
from a number of different, essentially random initial conditions.  Motivated
by these results though, one obvious question to ask would be: do there exist
symmetric solutions with 16 vortices, and/or asymmetric solutions with 18
vortices?  Various plausible initial conditions were therefore constructed,
but all attempts invariably equilibrated back to either of figures 8 or 9.

\section{Conclusion}

We have seen in this work how a magnetically ramped Taylor-Couette system
can yield qualitatively similar phase slip dynamics as in the more familiar
geometrically ramped system.  It would be of considerable interest to make
the comparison more quantitative, by deriving the equivalent of the Riecke
\& Paap \cite{RP87} phase-diffusion equation, and seeing whether it
can indeed explain the results presented here.  An analysis of the
numerically computed wave-numbers unfortunately proved inconclusive; the
results never vary by more than a few percent from the expected non-magnetic
values.  Note for example how all the bands in figures 3, 8 and 9 have
essentially the same axial extent, corresponding to round Taylor vortices.
It is thus not entirely clear why our solutions exhibit phase slip at all,
let alone two different types, occurring in different locations.  A weakly
nonlinear analysis as in other pattern-forming systems \cite{C05} might
also be fruitful.

Future numerical extensions of this work include a systematic search for
the spatiotemporal chaos predicted by Paap \& Riecke \cite{PR91}.
This will likely require imposing fields consisting of harmonics $\kappa$,
$2\kappa$, etc., analogous to the special choices of ramps needed in the Paap
\& Riecke formulation.  Also of interest is the magnetorotational instability
mentioned in the introduction, where the magnetic field has a destabilizing
rather than stabilizing influence; we would then expect to find Taylor
vortices where the field is strong rather than weak.

More generally, it is noteworthy that there do not appear to be any direct
numerical simulations of any of the geometrically ramped experiments mentioned
in the introduction.  The only related numerical work is \cite{S09}, who
did not address subcritical ramping though; instead they considered a scenario
where the radii of the cylinders vary along the length, but in such a way that
the onset of Taylor vortices still occurs simultaneously everywhere.  This
configuration yields vortices that form at one end and monotonically drift
toward the other, but without any bifurcations beyond the initial transition
from steady to periodic flows.

The most likely reason for this almost complete absence of numerical
work in geometrically ramped TC flows is that the geometry then becomes
too complicated to allow the very long integration times that are
required for equilibrated solutions to emerge.  In contrast, by keeping the
geometry simple, in this work we were able to exploit efficient
pseudo-spectral numerical methods that do allow very long integration times.
We suggest therefore that for numerical studies of general pattern formation
effects the magnetically ramped problem is more convenient.

Finally, it would be of great interest to see actual experiments done
on this problem.  Imposing magnetic fields as in Eq.\ (3) would be
straightforward, simply by a suitable array of periodically spaced external
Helmholtz coils.  The required field strengths are also easily achievable;
the existing PROMISE experiment \cite{S06,S14} already operates at slightly
larger Hartmann numbers.  The main difficulty would most likely be that the
Reynolds numbers here are considerably lower than in the PROMISE experiment,
where they are $O(10^3)$.  The entire flow velocities are then also reduced,
which could make them difficult to measure with the techniques commonly
used in liquid metal experiments of this type \cite{CEG13,PHKT}.  If these
difficulties could be overcome though, then magnetically modulated
Taylor-Couette flows would undoubtedly offer a rich variety of phenomena to
study experimentally.

\begin{acknowledgements}
RH was supported by STFC grant ST/K000853/1.  FK's visit to the
United Kingdom was supported by the Higher Education Commission of Pakistan.
\end{acknowledgements}

\end{document}